\documentclass[aps,pra,reprint,twocolumn,showpacs,floatfix,superscriptaddress]{revtex4-1}

\usepackage{amssymb,amsmath,amstext}                %%   American Physical Society math etc extensions
\usepackage{graphicx}                                               %%   Include figure files                                            
\usepackage{epstopdf}                                               %%   help with eps -> pdf 
\usepackage{color}                                                     %%   change color
\usepackage{bm}                                                        %%   bold math                                    
\usepackage{appendix}                                              %%   formating appendicies
\usepackage[utf8]{inputenc}
\usepackage{bbold}
\usepackage{bbm}
\usepackage{latexsym}
\usepackage{xcolor}
\usepackage{braket}
\definecolor{lblue} {RGB}{51,71,158}
\usepackage[colorlinks=true,citecolor=blue,linkcolor=blue,urlcolor=lblue]{hyperref}

\begin{document}

\title{Many-body localization transition in large quantum spin chains:
The mobility edge}
% Force line breaks with \\

\author{Titas Chanda}
\email{titas.chanda@uj.edu.pl}
\affiliation{Instytut Fizyki Teoretycznej, Uniwersytet Jagiello\'nski,  \L{}ojasiewicza 11, 30-348 Krak\'ow, Poland }

\author{Piotr Sierant}
\email{piotr.sierant@uj.edu.pl}
\affiliation{Instytut Fizyki Teoretycznej, Uniwersytet Jagiello\'nski,  \L{}ojasiewicza 11, 30-348 Krak\'ow, Poland }

\author{Jakub Zakrzewski}
\email{jakub.zakrzewski@uj.edu.pl}
\affiliation{Instytut Fizyki Teoretycznej, Uniwersytet Jagiello\'nski,  \L{}ojasiewicza 11, 30-348 Krak\'ow, Poland }
\affiliation{Mark Kac Complex
Systems Research Center, Uniwersytet Jagiello\'nski, Krak\'ow,
Poland. }

\date{\today}%% It is always \today, today,

%\pacs{03.67.Lx, 42.50.Dv}% PACS, the Physics and Astronomy

\begin{abstract}
Thermalization of random-field Heisenberg
spin chain is probed by time evolution of density correlation 
functions. Studying the impacts of  
average energies of initial product states on dynamics of the system,
we provide arguments in favor of the existence of a mobility edge
in the large system-size limit.
\end{abstract}

\maketitle

 {\it Introduction.--} Many-body localization (MBL)  \cite{Gornyi05, Basko06} is a robust mechanism that prevents reaching of 
 thermal equilibrium by quantum many-body systems \cite{Deutsch91, Srednicki94, Rigol08}.
 The phenomenon, originating from an
 interplay of interactions and disorder \cite{Nandkishore15, Alet18, Abanin19}, 
 has been studied numerically in various models: spin chains
 \cite{Santos04a, Oganesyan07,Pal10,Luitz15} that map onto spinless fermionic 
 chains, spinful fermions \cite{Mondaini15, Prelovsek16,Zakrzewski18,Kozarzewski18}
 or bosons \cite{Sierant17, Orell19, Hopjan19}.
 Despite those efforts, a complete understanding of the transition between ergodic
 and MBL phases is still lacking. While the recent works \cite{Goremykina19, Morningstar19, Dumitrescu19, Laflorencie20, Suntajs20}
 suggest a Kosterlitz-Thouless scaling at the MBL transition, it became clear that the exact diagonalization studies
 are subject to strong finite size effects \cite{Suntajs19, Sierant20b, Abanin19a} that 
 prevent one from reaching unambiguous conclusions about the thermodynamic limit \cite{Panda19, Sierant20c}.
 
 Alternatively, time evolution of large \cite{Doggen18} (or even infinite  \cite{Enss17})
 disordered many-body systems can be simulated with tensor network algorithms. Reaching large time 
 scales, necessary to assess thermalization properties \cite{Chanda19} is challenging especially in the vicinity of 
 transition to ergodic phase. Nevertheless, such an approach allows to obtain estimates for critical disorder strength
 for large system sizes \cite{Doggen18, Chanda19}, 
 in quasiperiodic systems \cite{Doggen19}, or even beyond one spatial dimension \cite{Hubig19, Doggen20}.
 An advantage of such an approach is that it directly mimics experimental observations of MBL 
 \cite{Schreiber15, Bordia17, Kohlert18, Lukin18, Rispoli18, Choi16, Smith16, Xu18}.

Typically, the transition between ergodic and MBL phases is induced by tuning 
 the disorder strength. Then the natural extension is -- can one envision a different
 control parameter? In this work, we consider energy as such a parameter. This immediately translates to a problem
 of the existence of \textit{many-body mobility edges}, i.e., energies that
 separate localized and extended states \cite{Basko06}. 
 The many-body mobility edges observed in early exact diagonalization studies of small systems  
 \cite{Kjall14,Luitz15,Mondragon15} were argued to be indistinguishable from finite-size effects in
 \cite{Roeck16}. The argument of \cite{Roeck16} is that local fluctuations in a system with a putative many-body mobility edge 
 can serve as mobile bubbles 
 inducing a global delocalization and hence no many-body mobility edge can exist.
The  existence of mobility edges is one of the fundamental problems of MBL,
 it leads to questions about phenomenology of systems with many-body mobility edge
 (since the description in terms of local integrals
 of motion \cite{Serbyn13b, Huse14, Ros15, Imbrie16, Wahl17, Mierzejewski18, Thomson18} does not apply in such 
 a scenario),
 or to the anomalous dynamics for a non-stationary initial state  due to energy fluctuations
\cite{Luitz-rev16,Luitz16b}.

% \old{An important property of MBL is the existence of \textit{many-body mobility edges}
% separating in energy localized from extended states \cite{Basko06}. This does not
% contradict the existence of zero conductivity of many body localized phase
% \cite{Kjall14} and is in contrast to single particle Anderson problem in uncorrelated 
% random potential \cite{Anderson58, Evers08} where all states are either localized or extended.
% Exact diagonalization studies \cite{Kjall14,Luitz15,Mondragon15} showing the mobility edge were challenged
% by \cite{Roeck16} who claimed that local fluctuations in the presence of a mobility edge would 
% induce a global delocalization. The numerical
%observations  presented in \cite{Kjall14,Luitz15,Mondragon15} were argued to be affected by finite-size effects.
%This calls for the numerical verification of the existence of many body mobility edges in large systems. Especially
%as mobility edge allows to induce the transition not only by changing the disorder amplitude but also by 
%changing the energy of the initial state (traditionally linked to its temperature). The presence of the mobility edge may lead to
%anomalous dynamics for such a non-stationary initial state (as due to energy fluctuations some of its
%parts may belong to localized and some to extended regime) \cite{Luitz-rev16,Luitz16b}.
     
\begin{figure}
\includegraphics[width=0.85\linewidth]{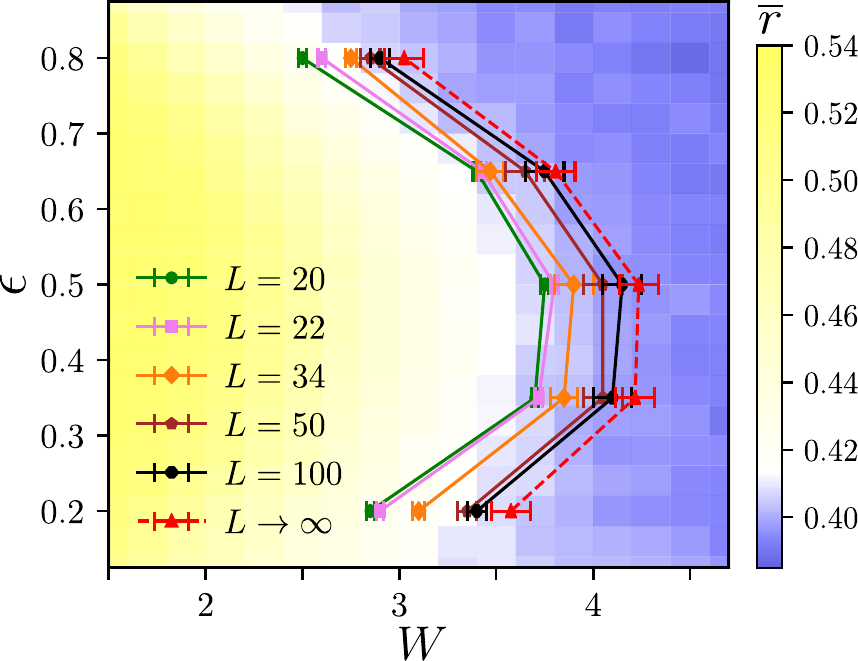}
  \caption{Phase diagram of random-field Heisenberg spin chain, disorder strength $W$
  on horizontal axis, rescaled energy $\epsilon$ on vertical axis. Background 
  shows color-coded value of average gap ratio $\overline r$ for system size $L=16$. 
  Solid lines show the position of boundary between ergodic and MBL phases 
  obtained in study of decay of density correlations in systems of size $L=20, 22, 26, 34, 50, 100$,
  dashed line shows results of extrapolation of the results to $L\rightarrow \infty$.
  }
  \label{edge14}
\end{figure}

  The aim of our work is to study many-body mobility edges at much larger system
  sizes than those available to exact diagonalization in an attempt to verify
  conclusions of  \cite{Roeck16}.  
  To that end, we employ {Chebyshev polynomial expansion of  the evolution operator \cite{Tal-Ezer84,Cheby91,Fehske08}} and
  the time-dependent variational principle (TDVP)  applied to matrix product states (MPS) \cite{Haegeman11, Koffel12, Haegeman16, Paeckel19}
  to simulate time dynamics of random-field Heisenberg spin chain. Our 
  approach is, in spirit, similar to that of \cite{Naldesi16} and \cite{Wei19} (and used for bosons in \cite{Yao20}). However, instead of 
  considering an injection of controllable amount of energy into ground state 
  of the system, we consider time evolution of initial product states 
  with specified average energies, exactly similar to what was done recently in spin quantum
  simulator \cite{Guo19}. Probing time decay of density correlation functions allows
  us to estimate the critical disorder strength as a function of energy of the initial state. 
  Studying systems of size up to $L=100$, we perform a finite size scaling of our results
  which provides arguments in favor of existence of mobility edge even in large systems.
  
%The paper is organized as follows: Sec. \ref{sec:model} briefly introduces the disordered model as well  numerical techniques considered in this study. We present our main results regarding the estimation of many-body mobility edge via the time evolution of large systems in Sec. \ref{sec:dyn} . Finally, we conclude in Sec. \ref{sec:conclu}.

%  \section{The model and method}
 % \label{sec:model}

{\it The model and methods.-- } We consider 1D random-field Heisenberg ($XXZ$) spin chain with the Hamiltonian given by
\begin{equation}
 H= J \sum_{i=1}^{L-1} \left( S^x_{i}S^x_{i+1}+S^y_{i}S^y_{i+1} + S^z_{i}S^z_{i+1}  \right) + \sum_{i=1}^{L} h_i S^z_i,
 \label{eq:XXZ}
\end{equation}
where  $S^{\alpha}_i, \ \alpha = x, y, z,$ are spin-1/2 matrices, $J=1$ is fixed to be the unit of energy,
and $h_i \in [-W, W]$ 
are independent, uniformly distributed random variables. In this work, we consider open boundary conditions in the Hamiltonian \eqref{eq:XXZ}. 
 The random-field Heisenberg spin chain has been
widely studied in the MBL context
\cite{Berkelbach10, Luitz15, Agarwal15, Bera15, Enss17, Bera17, Herviou19, Colmenarez19, Sierant20}, which has made it the \textit{de facto} standard model of MBL studies.

The transition between ergodic and MBL phases is reflected in change of 
statistical properties of energy levels of the system. A common approach is 
to consider the gap ratio 
$r_i=\frac{\min \{E_{i+2}-E_{i+1},E_{i+1}-E_{i} \} }{\max\{E_{i+2}-E_{i+1},E_{i+1}-E_{i}\}} $, 
where $E_i$ are the energy eigenvalues of the system. Averaging the gap ratio over part of the spectrum 
of the system and over disorder realizations, one obtains an average gap ratio $\overline r$, which 
 differentiates between level statistics of 
ergodic system \cite{Oganesyan07, Atas13}, well described by Gaussian orthogonal ensemble of random matrices, 
for which $\overline r \approx 0.53$ and between Poissonian statistics of eigenvalues in MBL phase (for which 
$\overline r \approx 0.39$). The later
arises due to emergent integrability resulting from the presence of local
integrals of motion \cite{Serbyn13b, Huse14, Ros15, Imbrie16, Wahl17, Mierzejewski18, Thomson18}.

To reveal the dependence of ergodic-MBL transition on energy, 
the gap ratios $r_i$ are averaged over only a certain number of eigenvalues with energies
close to a rescaled energy $\epsilon=(E-E_{\min})/(E_{\max}-E_{\min})$, where $E_{\min}$ ($E_{\max}$)
is the energy of the ground (highest excited) state. Such a
calculation of average gap ratio (supported with results for other probes of localization) 
for random-field Heisenberg spin chain reveals that the ergodic region has shape
of a characteristic lobe on the phase diagram in variables of the 
rescaled energy $\epsilon$ and disorder strength $W$ \cite{Luitz15} . The average gap ratio, obtained in exact 
diagonalization of random field Heisenberg spin chain of size $L=16$, is plotted as a function
of $\epsilon$ and $W$ in the background of Fig.~\ref{edge14}.

To probe the transition between ergodic and MBL phases with time evolution, we propose  the following
protocol. We consider an initial state $| \psi \rangle= | \sigma_1, \ldots, \sigma_L \rangle$, where 
$\sigma_i=\uparrow,\downarrow$ are chosen randomly with constraint that 
the average rescaled energy 
$\epsilon_{\psi} = (\langle \psi | H | \psi \rangle-E_{\min})/(E_{\max}-E_{\min})$ of this state 
lies withing the range $[\epsilon - \delta \epsilon, \epsilon + \delta \epsilon]$
corresponding to a given rescaled energy
$\epsilon$, where $\delta \epsilon$ is a small tolerance (we take $\delta \epsilon =0.01$).
To calculate $\epsilon_{\psi}$ for $L\leqslant 26$ we find $E_{\max}, E_{\min}$
with the standard Lanczos algorithm \cite{Lanczos50}. For larger system sizes, $E_{\min}$ and $E_{\max}$ 
are calculated using  density matrix renormalization group (DMRG) algorithm 
\cite{white_prl_1992, white_prb_1993, schollwock_rmp_2005, schollwock_aop_2011, Orus14}(see  \cite{supple} for details).

Subsequently, we calculate time evolved state $| \psi(t) \rangle= e^{-iHt}| \psi \rangle$ with the standard
Chebyshev expansion of the evolution operator \cite{Fehske08} for $L \leqslant 26$. For larger system
sizes, we use the recently developed TDVP algorithm \cite{Haegeman11, Koffel12, Haegeman16, Paeckel19}.
Technically, we follow \cite{Chanda19, Chanda20} and employ a hybrid of two-site and one-site versions 
of TDVP \cite{Paeckel19, Goto19} (see  \cite{supple} for details).

\begin{figure}
\includegraphics[width=\linewidth]{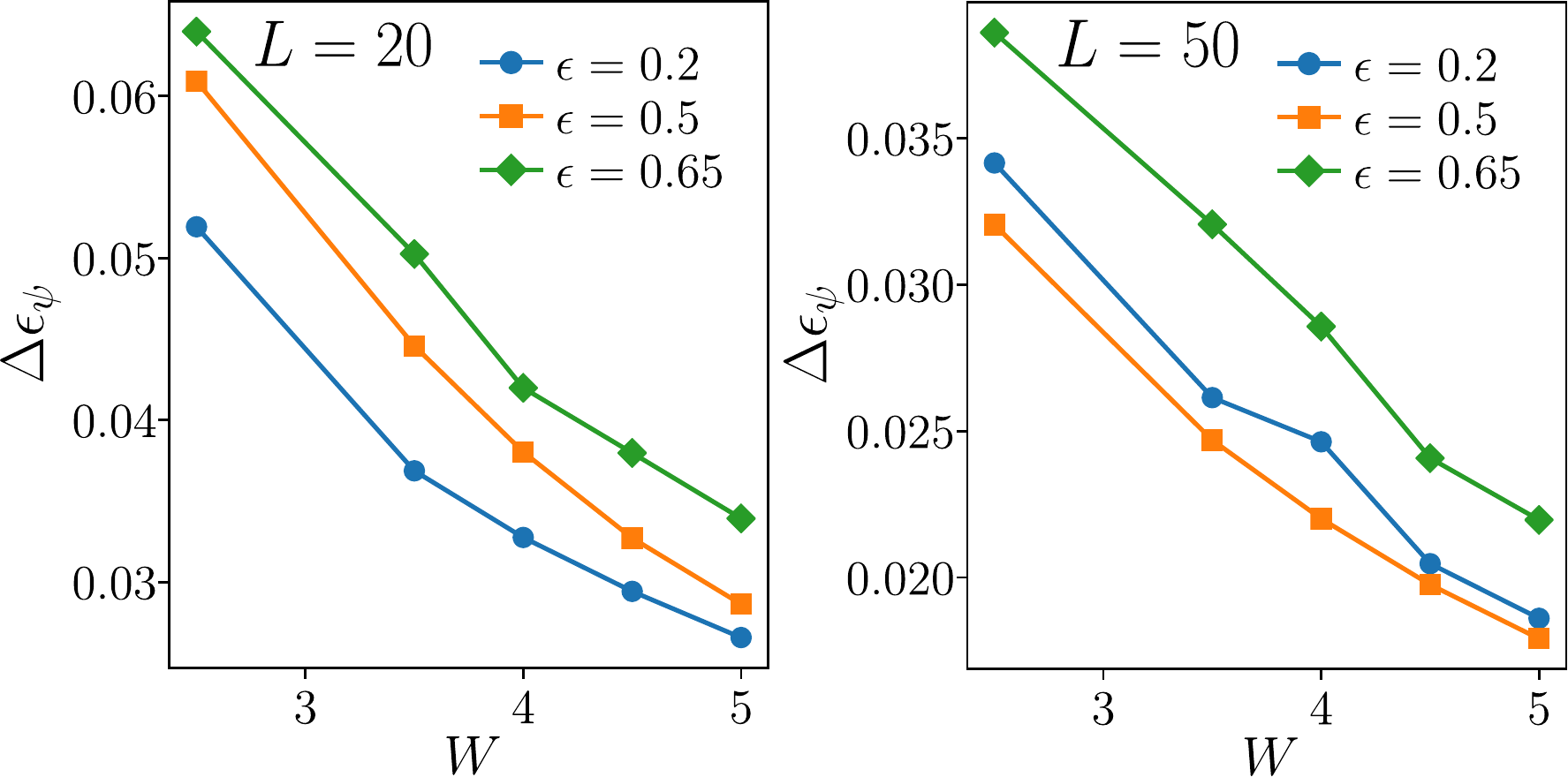}
  \caption{Disorder averaged standard deviation $\Delta \epsilon_{\psi}$ of rescaled energy of the initial states as a function of disorder
  strength $W$ for three exemplary rescaled energies $\epsilon$. Left: system size $L=20$, Right: system size: $L=50$.
  }
  \label{stdE}
\end{figure}

Our quantity of interest is the density correlation function
\begin{equation}
 \label{eqcor1}
 C(t) = D \sum_{i=1+l_0}^{L-l_0}  \langle \psi(t) | S^z_i|\psi(t) \rangle   \langle \psi | S^z_i |\psi \rangle,
\end{equation}
where the constant $D$ assures that $C(0)=1$ and $l_0>0$ diminishes the influence of boundaries (in our calculations,
we take $l_0=2$). The standard deviation of the rescaled energy 
\begin{equation}
\Delta \epsilon_{\psi}= \left( \langle \psi | \left( (H-E_{\min})/(E_{\max}-E_{\min}) - \epsilon_{\psi}\right)^2 | \psi \rangle\right)^{1/2}  
\end{equation}
is smaller than $0.1$ for disorder strengths that we consider in this work as shown in Fig.~\ref{stdE}. Those relatively small
fluctuations of energy suggest that the properties of eigenstates at the rescaled energy $\epsilon$
can be well probed by time evolution of the state $\ket{\psi}$ and reflected, in particular, by the density correlation function 
$C(t)$.

%\section{Quench dynamics}
%\label{sec:dyn}
%
%\subsection{Disorder strength dependence}

\begin{figure}
\includegraphics[width=0.7\linewidth]{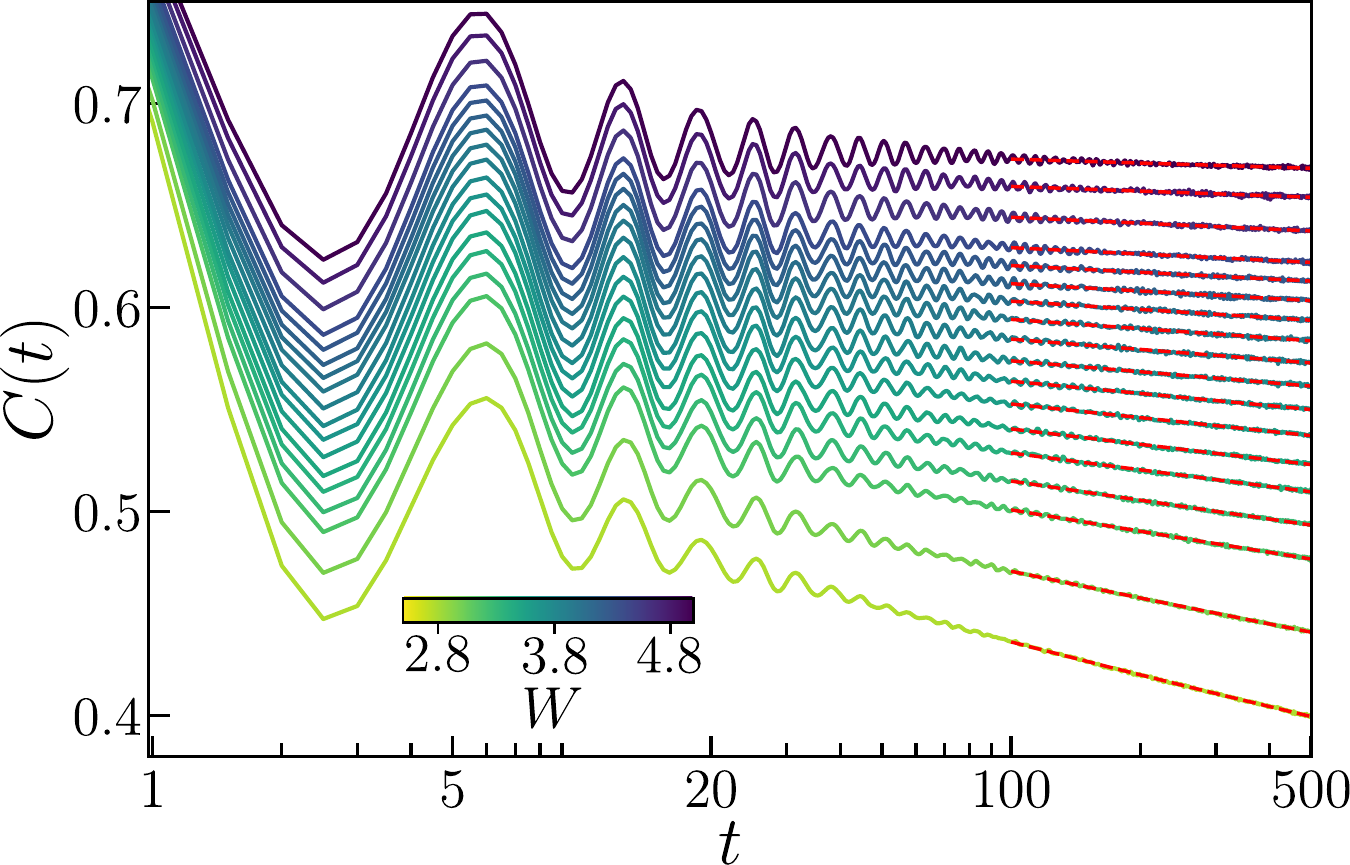}
  \caption{Quench dynamics in disordered $XXZ$ spin chain.
  Density correlation function $C(t)$ for system size $L=20$ and various disorder strengths
  $W=2.8, ... 5$ (color coded) averaged over $10000$ disorder realizations, rescaled energy of the initial state $\epsilon=0.5$, 
  power-law fits $C(t)\propto t^{-\beta}$ for $t \in[100, 500]$ are denoted by the
  dashed lines.}
  \label{figCt}
\end{figure}

{\it Quench dynamics: dependence on disorder strength.--} Fig.~\ref{figCt} shows the density correlation functions $C(t)$ obtained for the random-field Heisenberg spin chain
of a fixed size $L=20$, for rescaled energy $\epsilon=0.5$ of the initial state. 
The correlation function decreases in time, with some oscillations superimposed \cite{Luitz16}.
For small disorder strength, e.g. $W=2.8$ the eigenstate thermalization hypothesis \cite{Rigol08, Alessio16}
is valid for the system, and in the long time limit 
the correlation function vanishes $C(t) \stackrel{t\rightarrow \infty }{\rightarrow}0$ as system loses the memory of the
initial state. In contrast, for large disorder strength, e.g., $W=5$, a non-zero stationary value of the correlation
function $C(t) \stackrel{t\rightarrow \infty }{\rightarrow} c_0 > 0$ is admitted showing that the system is non-ergodic.
The first experimental signatures of MBL were obtained in 
study of time evolution of \textit{imbalance} \cite{Schreiber15}, 
quantity analogous to the density correlation function -- for quantitative comparison 
of the two quantities see \cite{supple}. 

At large times ($t>100$), the decay of the correlation function is well described by a power law,
$C(t) \propto t^{-\beta}$. Griffiths rare regions are one possible explanation of this behavior
\cite{Agarwal16}. However, it was shown experimentally and numerically that time dynamics in quasiperiodic
potentials, where Griffiths regions are necessarily absent, 
have analogous features \cite{Luschen17, Bera17, Weiner19}.
Regardless of the origin of the power law decay of the correlation function, the disorder strength dependence of the
exponent $\beta$ can be used to locate the onset of ergodicity breaking in the system.

\begin{figure}
\includegraphics[width=0.75\linewidth]{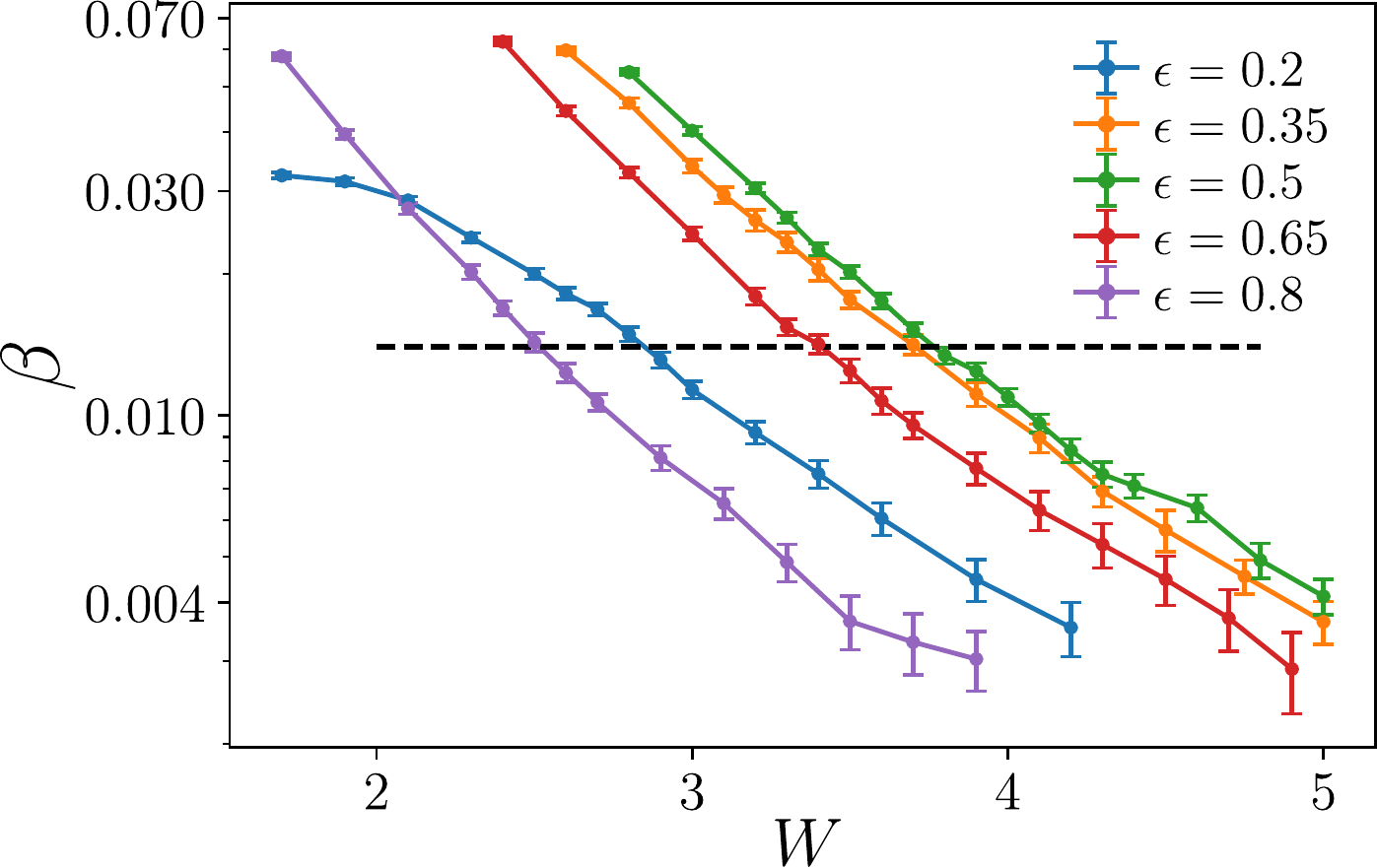}
  \caption{The exponent $\beta$, obtained in fitting the density correlation
  function $C(t)$ with an algebraic decay $a_0 t^{-\beta}$ in interval $t\in[100,500]$, is
  plotted as function of disorder strength $W$. The errorbars represent the $1\sigma$ errors of the fitting obtained from statistical resampling of disorder
  realizations.
  The 
  system size is $L=20$, results for various rescaled energies $\epsilon$ of the initial state
  are shown. The dashed line shows the cut-off exponent $\beta_0 = 0.014$.
  }
   \label{betaL20}
\end{figure}  

The exponent $\beta$ governing the decay of the density correlation function is shown in 
Fig.~\ref{betaL20}(a). Let us first concentrate on the results in the middle of the spectrum
($\epsilon=0.5$). In the considered interval of disorder strength $W$, 
the exponent decreases exponentially with $W$ with a good approximation 
$\beta \propto e^{-W/\Omega}$. The large number of disorder realizations (10000) used in 
calculation of $C(t)$ allows us to see that even at the large disorder strength $W=5$
the exponent $\beta = 4.1(4)\cdot 10^{-3}$ is non-vanishing.
If the power-law decay $C(t)=a_0 t^{-\beta}$ prevailed for $t\rightarrow \infty$, 
the density correlation function would vanish in the long-time limit and the system would be ergodic.
This, however,  does not happen for $L=20$, as after the so-called Heisenberg time
$t_H$ discreteness of spectrum manifests itself in 
saturation of the correlation functions \cite{Torres-Herrera15, TorresHerrera17, Torres-Herrera18, Schiulaz19},
so that one would observe $C(t) \stackrel{t\rightarrow \infty }{\rightarrow} c_0 > 0$ for $W=5$ and $L=20$. 
The Heisenberg time $t_H$  increases exponentially with the system size $L$. 
This illustrates a difficulty in locating the MBL transition using time dynamics 
of large systems on time scales of few hundred $J^{-1}$ 
accessible to tensor network methods (or to current experiments with e.g., ultra-cold
atoms): one cannot predict whether a slow decay of correlation functions governed 
by an exponent $\beta \ll 1$ observed, for example $t\in[100,500]$, 
will eventually lead to $C(t) \stackrel{t\rightarrow \infty }{\rightarrow} 0$ or not.

To resolve the difficulties, the work of \cite{Doggen18} assumes that the value of the exponent 
$\beta$ must be vanishing within error bars to be 
compatible with saturation of correlation functions in the long time limit.
The drawback of this criterion is that the error bar of $\beta$ depends on the 
number of disorder realizations used in calculation of the correlation function. 
Therefore, we introduce a cut-off $\beta_0$: disorder strength $W_C(L)$ 
for which $\beta=\beta_0$ is regarded as disorder strength for transition
to MBL phase at system size $L$. 
Exact diagonalization results show that: i) collapse of data for $L\leqslant 22$
gives a critical disorder strength $W_C\approx3.7$; ii) the  similar values 
$W_C\approx 3.8$ or $W_C\approx 4.2$ are obtained in asymmetric scaling on the 
two sides of the transition; iii) the breakdown of the volume-law scaling of
entanglement entropy gives an estimate $W_C=3.75$ at system size $L=20$ \cite{Sierant20c}.
The obtained results for $\beta$ at $L=20$ and $\epsilon=0.5$ show that the cut-off value $\beta_0=0.014$ 
is consistent with the above estimates for the critical disorder strength obtained from exact diagonalizations, see Fig. \ref{betaL20}. 
The assumed cut-off value $\beta_0$ is 
nearly independent of the target energy and system size (for further details see
\cite{supple}).
Consequently, throughout this work, we use $\beta_0=0.014$ as a threshold value which separates 
ergodic and MBL regimes for all system sizes and energies of the initial state we consider.

The values of $\beta$ presented in Fig.~\ref{betaL20}(a) show that the increase of
disorder strength $W$ slows down the dynamics more severely for rescaled energies of initial state
different than $\epsilon=0.5$. Notably, the exponent $\beta$ decreases exponentially with $W$:
$\beta \propto e^{-W/\Omega}$ (where $\Omega$ is a constant) in a wide regime of
disorder strengths. This resembles the scaling of Thouless time 
$t_{Th}\propto \mathrm{e}^{-W/W_0 }$ observed in exact diagonalization data in \cite{Suntajs19}.

%\subsection{System-size dependence}

\begin{figure}
\includegraphics[width=\linewidth]{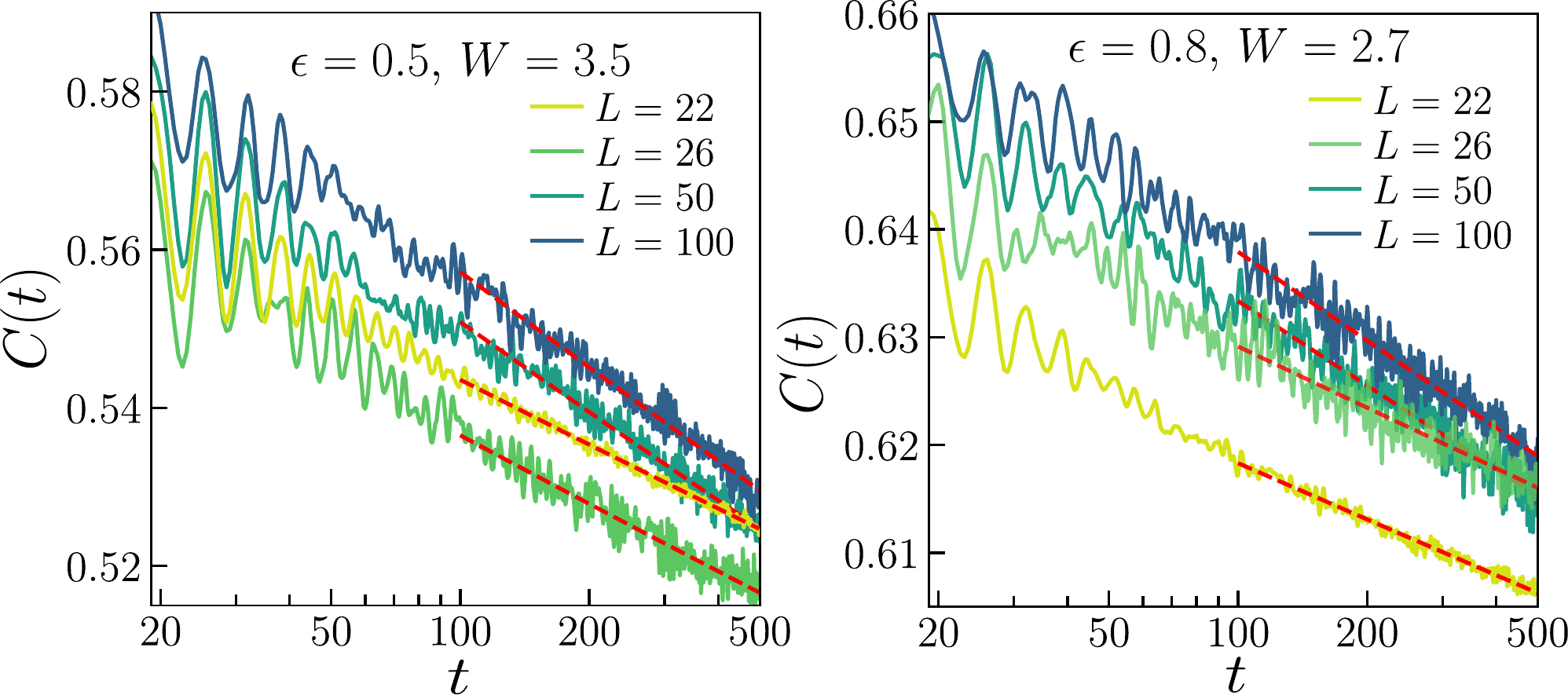}
  \caption{Time evolution of density correlation function $C(t)$ for rescaled energy of initial state
  $\epsilon=0.5$ ($\epsilon=0.8$) and disorder strength $W=3.5$ ($W=2.7$) in panel right (left). The system
  size $L$ varies from $22$ to $100$. The dashed lines denote power-law fits $C(t)=a_0 t^{-\beta}$
  in the $t\in[100,500]$ interval. 
  }
  \label{figTevolL}
\end{figure}

\begin{figure}
\includegraphics[width=\linewidth]{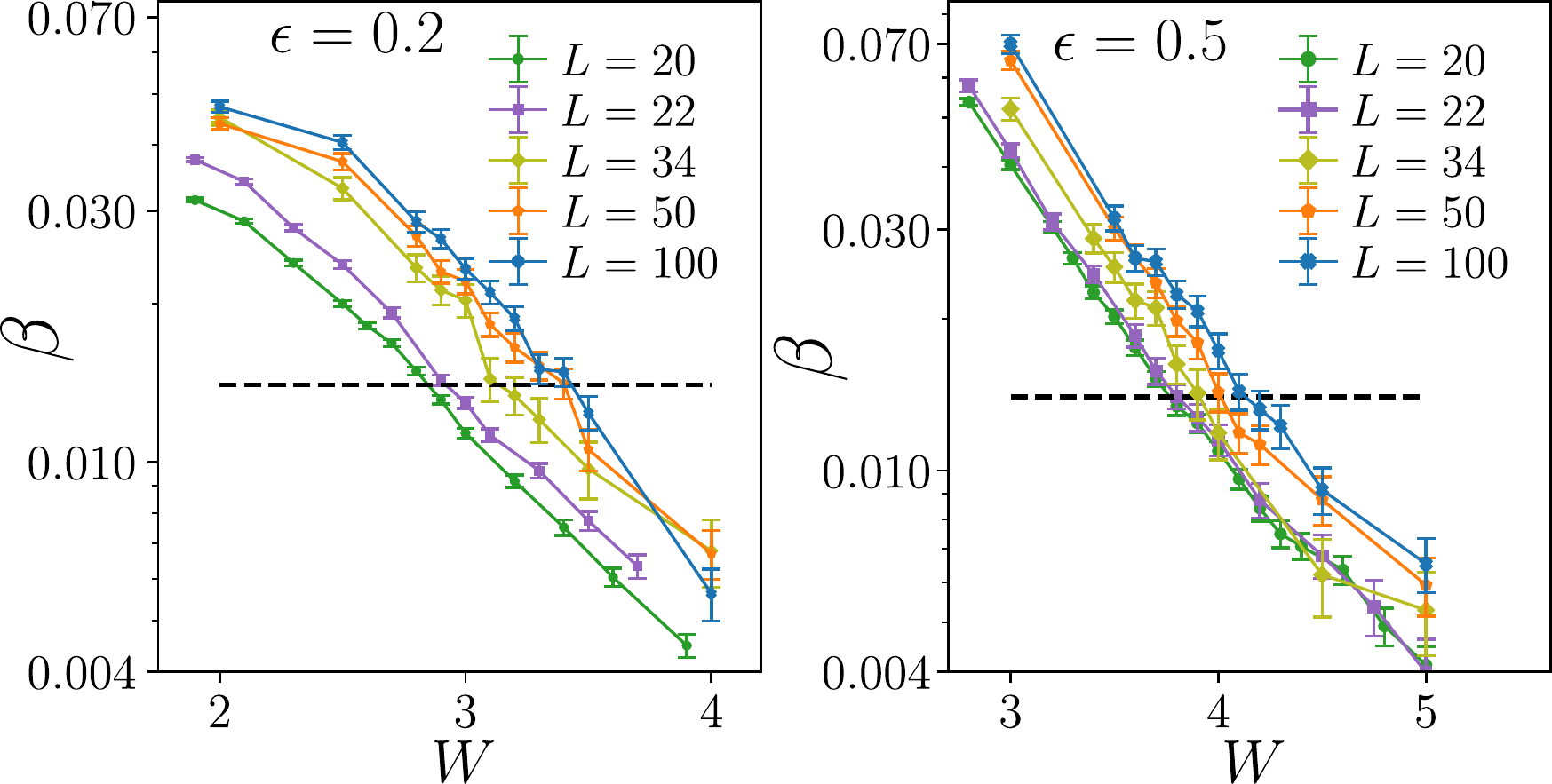}
  \caption{The exponent $\beta$ obtained in fitting the density correlation
  function $C(t)$ with an algebraic decay $a_0 t^{-\beta}$ in interval $t\in[100,500]$
  for the rescaled energy  $\epsilon=0.2$ ($\epsilon=0.5$) of the initial state
  shown in the left (right) panel. Data shown for system sizes $L=20,22, 34, 50, 100$.
  The error bars show $1\sigma$ errors of $\beta$ obtained in resampling of disorder
  realizations. The dashed lines show the cut-off exponent $\beta_0$.
  }
  \label{figExpL}
\end{figure}

{\it Quench dynamics: dependence on system size.--} Density correlation function $C(t)$ for larger system sizes are shown for two 
exemplary pairs of disorder strength $W$ and initial rescaled energy $\epsilon$ in Fig.~\ref{figTevolL}. 
The decay of $C(t)$ at large times is well fitted by an algebraic dependence $C(t) \propto t^{-\beta}$.
The exponents $\beta$ obtained in the fitting of power-law decay to $C(t)$ 
are shown for two exemplary values of the rescaled energy $\epsilon$ of the initial states in
Fig.~\ref{figExpL}.
For a given disorder strength $W$, we observe a clear increase of $\beta$ with increasing system size.
Interestingly, the shift is, to a good approximation, uniform for all disorder strengths so that the 
exponential decrease $\beta \propto e^{-W/\Omega}$ (at sufficiently large $W$) is observed for all 
considered system sizes. Let us mention here that  we consider 400 realizations of
disorder for $L=34, 50$, and 200 realizations for 
$L=100$ for each values of $\epsilon$ and $W.$ For small system sizes ($L=20, 22, 26$)
we consider between $10000$ and $500$ disorder realizations.

\begin{figure}
\includegraphics[width=0.99\linewidth]{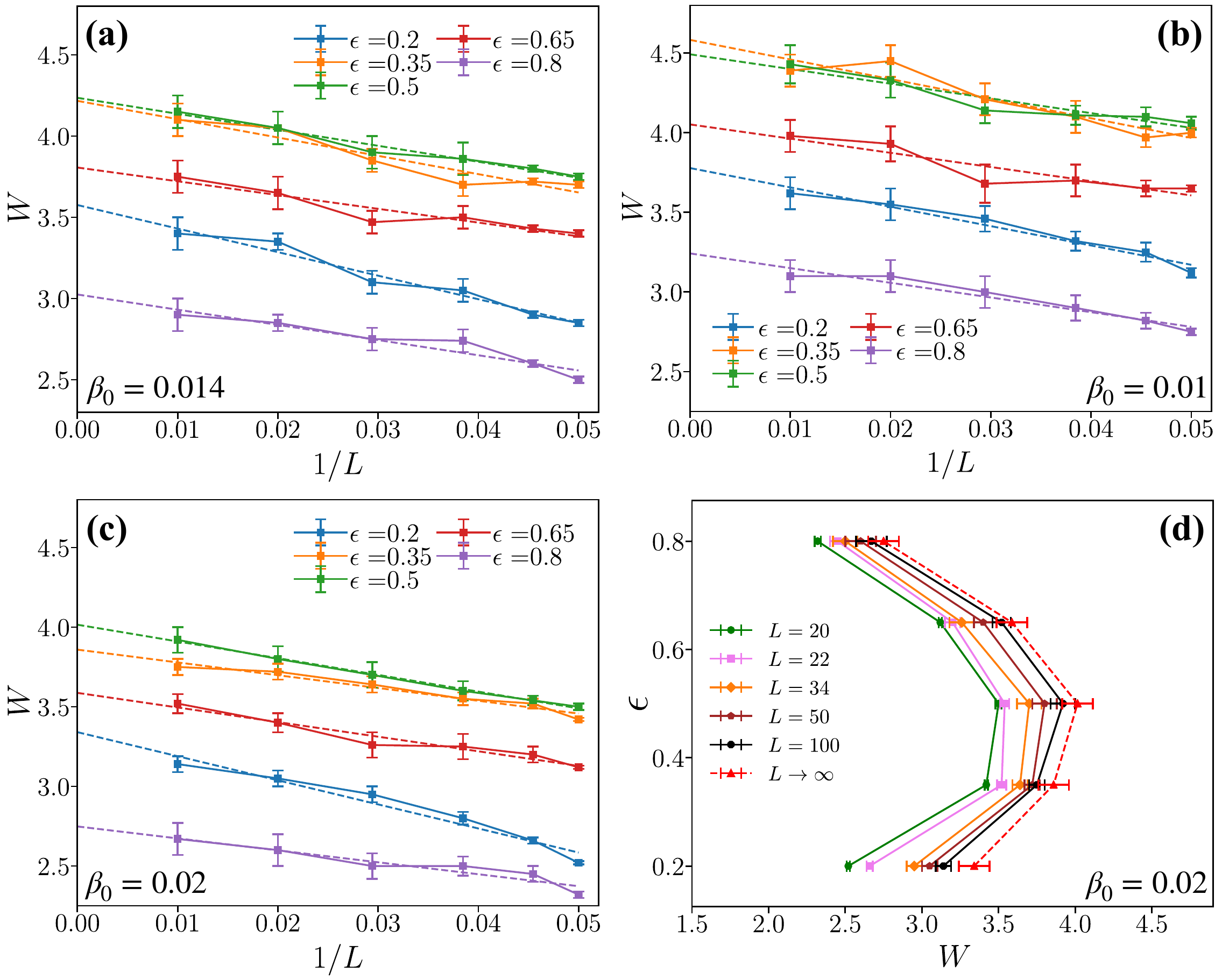}
  \caption{(a) Disorder strength $W_C(L)$ for which decay of correlation function is governed by
  power-law with $\beta=\beta_0$ plotted as function of $1/L$ where $L$ is the system size.
  Results shown for various rescaled energies $\epsilon$ of initial state. Available data are fitted with linear
  functions $W_C(1/L) = a/L +W_C(\infty)$ which allow extrapolation to $L\rightarrow \infty$.
(b)-(c) Same as in (a), but with two different choices of cut-off $\beta_0$ (0.01 and 0.02 respectively).
   (d) The shape of the edge between localized and delocalized regions in $(\epsilon, W)$-plane for different system-sizes $L = 20, 20, 34, 50, 100$ obtained with threshold $\beta_0=0.02$. Dashed line shows the extrapolation of the results for $L \rightarrow \infty$.
  }
  \label{diso_1L}
\end{figure}

We obtain estimates for  disorder strength $W_C(L)$ for transition to MBL phase 
by finding the crossings of $\beta(W)$ curve for given system size $L$ with the $\beta=\beta_0$ line.
Results of this procedure are shown in Fig.~\ref{diso_1L}(a). 
The disorder strength $W_C(L)$ depend, within the estimated error bars, linearly on the inverse of the system size $L$ with clear
growth of $W_C(L)$ as the system size increases. On one hand, this trend allows us, by means of a linear fit $W_C(1/L) = A/L +W_C(\infty)$,
to extrapolate the results to $L\rightarrow \infty$ and to obtain the estimate of critical disorder strength $W_C(\infty)$ for transition 
to MBL phase. 

On the other hand, we observe that the slopes $A$ are similar for all of the considered rescaled energies of the
initial state. Thus, the shape of the boundary between ergodic and MBL regimes observed for $L=20$ does not change 
considerably when the system size is increased. This is visible in Fig.~\ref{edge14}. The points for various system sizes $L$
are precisely the values of $W_C(L)$ obtained from the condition $\beta=\beta_0$. The characteristic shape of the lobe does not 
change when the system sizes increases from $L=20$ to $L=100$ and is preserved even after the extrapolation to $L\rightarrow\infty$.
Therefore, there exists a certain range of disorder strengths such that the states for $\epsilon<\epsilon^L_{ME}$ 
are localized, states for $\epsilon^L_{ME}< \epsilon<\epsilon^U_{ME}$ are extended and states for $ \epsilon> \epsilon^U_{ME}$
are again localized. Thus, our results indicate that the system indeed possesses a many-body mobility edge in the thermodynamic limit.

Upto now, the results are reported with the threshold value $\beta_0 = 0.014$. 
However, the qualitative results and the conclusion about the existence of mobility
edge in large systems remain unaltered for different choices of $\beta_0$, which we show  
in Figs. \ref{diso_1L}(b) and (c) by considering $\beta_0 = 0.01$ and $0.02$ respectively.
However, very small choice of $\beta_0$ (e.g., 0.01) results in larger error bars, 
which points towards the difficulty of obtaining the saturation of the correlation 
function within finite interval of time with finite number of disorder realizations.
Fig. \ref{diso_1L}(d) shows the shapes of the boundary between MBL and delocalized 
obtained for $\beta_0 = 0.02$ at different system-sizes, which remain qualitatively 
same as those for $\beta_0 = 0.014$.

%\section{Discussion and outlook}
%\label{sec:conclu}

{\it Discussion and outlook.--} Chebyshev polynomial expansion of  the time evolution operator and the
TDVP method applied to MPS allowed us to study the problem of energy dependence of 
the transition between ergodic and MBL phases in large disordered quantum spin chains. 
Introducing a cut-off value of exponent $\beta$ of power-law decay in time of density correlation function,
we were able to probe the transition for different rescaled energies of the initial state. For small system-sizes
(e.g., $L=20$), our approach gives results consistent with exact diagonalization. Importantly, our method allows to 
consider much larger system sizes ($L=100$) for which it predicts an existence of a mobility edge. 

The disorder strength $W_C(L)$ is a lower bound on the transition to MBL phase: 
the residual decay of density correlations with exponent $\beta_0$ is insufficient 
to restore the uniform density profile for system size $L=20$, but it is possible that it leads to 
an eventual decay of correlation function for larger system sizes.

The protocol we considered is, in principle,
experimentally realizable. Our results can be verified experimentally if the setup
of \cite{Guo19} was scaled to larger system sizes.
Many-body mobility edge arises also in disordered Bose-Hubbard models \cite{Sierant18}. It can be probed 
by a quench protocol analogous to the one considered in this work. Since the bosonic models 
allow for occupations in each site larger than unity, density wave-like states that are easier to obtain in an experiment with 
ultra-cold atoms can be use to probe the 
many-body mobility edge \cite{Sierant18, Yao20}. When this work was close to completion, we have learnt about a recent study \cite{Brighi20} where many-body mobility edge with respect to particle numbers were shown to exist in a correlated hopping model of hardcore bosons.

\acknowledgments
 Support of the Polish National Science Centre via
grants Unisono 2017/25/Z/ST2/03029 (T.C.) (under QTFLAG Quantera collaboration), Opus 2015/19/B/ST2/01028 (P.S.),  and Opus 2019/35/B/ST2/00034 (J.Z.)  is acknowledged.
P.S. thanks the Polish National Science Centre for an additional support via Etiuda programme 2018/28/T/ST2/00401 as well as the Foundation for Polish Science (FNP) through scholarship START. 
The partial support by  PL-Grid Infrastructure is also acknowledged. The MPS-based techniques have been implemented using
ITensor library v2 (\url{https://itensor.org}).

\bibliographystyle{apsrev4-1}
%\bibliography{ref_05_20.bib} 
\bibliography{mobil_v3psAtc.bbl} 

\end{document}

\title{Supplementary material to \\
``Many-body localization transition in large quantum spin chains:
The mobility edge''}
% Force line breaks with \\

\author{Titas Chanda}
\email{titas.chanda@uj.edu.pl}
\affiliation{Instytut Fizyki Teoretycznej, Uniwersytet Jagiello\'nski,  \L{}ojasiewicza 11, 30-348 Krak\'ow, Poland }

\author{Piotr Sierant}
\email{piotr.sierant@uj.edu.pl}
\affiliation{Instytut Fizyki Teoretycznej, Uniwersytet Jagiello\'nski,  \L{}ojasiewicza 11, 30-348 Krak\'ow, Poland }

\author{Jakub Zakrzewski}
\email{jakub.zakrzewski@uj.edu.pl}
\affiliation{Instytut Fizyki Teoretycznej, Uniwersytet Jagiello\'nski,  \L{}ojasiewicza 11, 30-348 Krak\'ow, Poland }
\affiliation{Mark Kac Complex
Systems Research Center, Uniwersytet Jagiello\'nski, Krak\'ow,
Poland. }

\date{\today}%% It is always \today, today,

%\pacs{03.67.Lx, 42.50.Dv}% PACS, the Physics and Astronomy

\begin{abstract}
Here, we first briefly describe the computational methods performed via matrix product state (MPS) in the main text. Then we also supplement the main text via additional results. Specifically, we discuss the dependence of the cut-off $\beta_0$ on the energy density and system-size.
\end{abstract}

\maketitle

\subsection{Brief descriptions of the MPS calculations} 
\label{app2}

For large system sizes, i.e., $L > 26$, we use matrix product states (MPS) ansatz \cite{Orus14, schollwock_aop_2011} 
to represent the quantum state of the disordered Heisenberg chain.
The energies of the ground state ($E_{\min}$) and highest excited state ($E_{\max}$) are obtained using standard
density matrix renormalization group method (DMRG) 
\cite{white_prl_1992, white_prb_1993, schollwock_rmp_2005, schollwock_aop_2011, Orus14} .
$E_{\min}$ can simply be found from the ground-state search of $H$ of Eq. (1) of the main text 
using DMRG, while that of $-H$ gives us $-E_{\max}$. In these calculations, we keep
truncation error due to singular value decomposition (SVD) below $10^{-12}$, i.e., we 
discard states corresponding to smallest singular values $\lambda_n$ that satisfy 
$\frac{\sum_{n \in \textrm{discarded}} \lambda_n^2}{\sum_{n \in \textrm{all}} \lambda^2_n} < 10^{-12}$.

For time evolution large systems, i.e., $L > 26$, we employ MPS based time-dependent variational principle (TDVP) method \cite{Haegeman11, Koffel12, Haegeman16, Paeckel19}.
Specifically, we use `hybrid' scheme of TDVP \cite{Paeckel19, Goto19, Chanda19, Chanda20} with 
step-size $\delta t = 0.1$, where two-site variant of TDVP is used initially to dynamically grow 
the bond dimension of the MPS, until the bond dimension in the bulk of the MPS is saturated to a 
predetermined value $\chi_{\max}$. One-site variant of TDVP is employed in the subsequent dynamics.
In our calculations, we fix $\chi_{\max}$ to be $384$, which gives reliable results up to $t=500$. 
For a more detailed analysis of the convergence of TDVP in the disordered Heisenberg chain, see \cite{Chanda19}.

 \subsection{Imbalance vs density correlation function}
 \label{app1}
 
 The N{\'e}el state
 \begin{equation}
 | \psi_{\textrm{N{\'e}el}} \rangle= | \uparrow, \downarrow, \uparrow, \ldots \rangle  
 \end{equation}
was used in previous works \cite{Doggen18,Chanda19} to locate the transition to MBL phase.
It is worth to mention that density wave states, analogous to the N{\'e}el state
were used in  experiment with ultra-cold fermions \cite{Schreiber15} that demonstrated
signatures of MBL transition.

\begin{figure}
\includegraphics[width=0.98\linewidth]{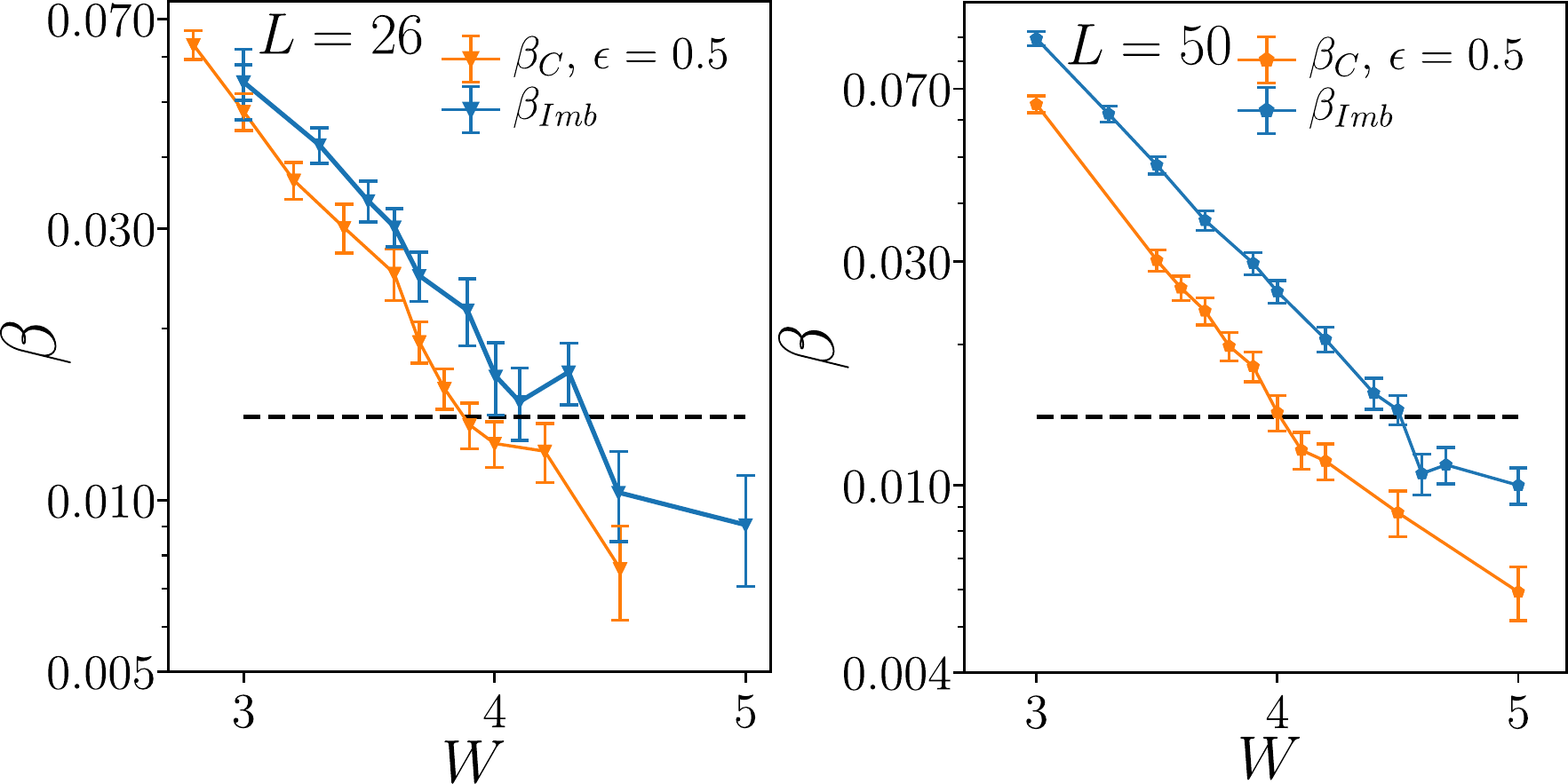}
  \caption{The exponent $\beta_C$ obtained in fitting the density correlation
  function $C(t)$ with an algebraic decay $a_0 t^{-\beta}$ in interval $t\in[100,500]$
  for rescaled energy  $\epsilon=0.5$ of the initial state together with exponent $\beta_{Imb}$
  obtained from the fitting of $I(t)$ for N\'eel state in the same time interval. Error bars dictate $1\sigma$ errors in $\beta$ obtained from statistical resampling
  procedure. 
  Left: system size $L=26$, Right: system size $L=50$.
  }
  \label{imbs}
\end{figure}

Imbalance $I(t)$ is the common name for the density correlation function (Eq. (2) of the main text) calculated for the
N{\'e}el state. The imbalance behaves similarly to the density correlation function: initially it decreases and oscillates,
subsequently it decays algebraically in time. Fig.~\ref{imbs} shows comparison of exponents $\beta_C$ and $\beta_{Imb}$
that govern the power-law decay of $C(t)$ correlation function for rescaled energy of initial state $\epsilon=0.5$
(which is the rescaled energy corresponding to the N{\'e}el state),
and power-law decay of imbalance of N{\'e}el state. The exponents $\beta_{Imb}$ are consistently  bigger than 
$\beta_C$. This fact may be traced back to the number of states coupled by the Hamiltonian (Eq. (1) of the main text)
to state $| \psi_{\textrm{N{\'e}el}} \rangle$ which is higher than the average number of states coupled to a typical Fock state 
with $\epsilon=0.5$. The absence of aligned spins on neighboring sites makes the thermalization process more efficient for N{\'e}el state.
The inequality $\beta_C<\beta_{Imb}$ justifies the cut-off beta value $\beta^{Imb}_0=0.02 $ for decay of imbalance in 
N{\'e}el state assumed in \cite{Chanda19}.

 \begin{figure}
\includegraphics[width=0.98\linewidth]{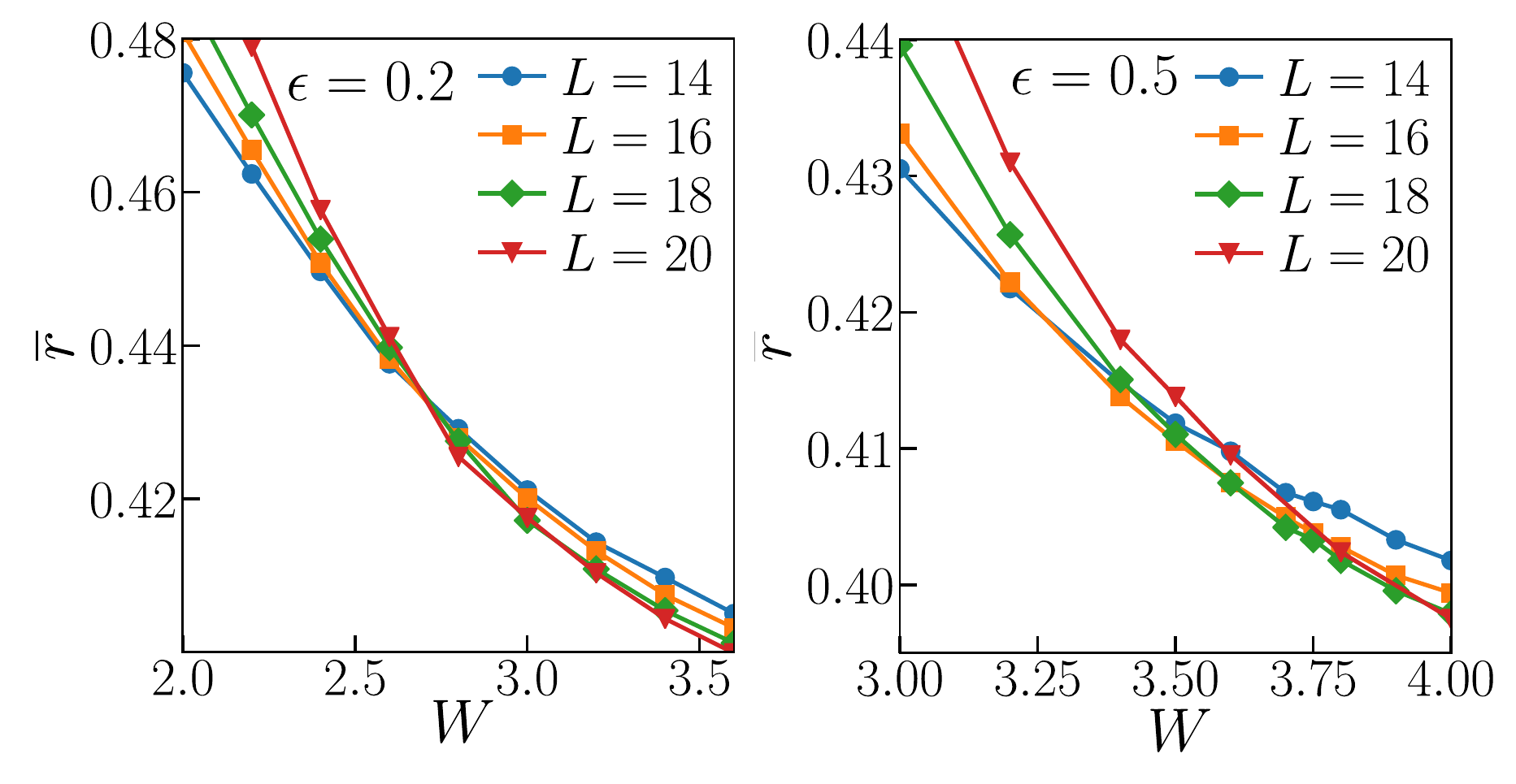}
  \caption{The average gap ratio $\overline r$ for random field
  Heisenberg spin chain for rescaled energies $\epsilon=0.2$ (left panel) and
  $\epsilon=0.5$ (right panel) for different system-sizes $L \leq 20$.
  }
  \label{rbars0205}
\end{figure}  

 \subsection{Energy and system size dependence of the cut-off value $\beta_0$ }
 \label{app3}
 The cut-off value $\beta_0=0.014$ assumed in the main text, was obtained from comparison of exact diagonalization
 results for systems of size $L\leqslant 22$ with the exponent $\beta$ governing the power-law decay of density correlation
 function $C(t)$ for system size $L=20$. This procedure leads, however, to questions about dependence of the cut-off $\beta_0$ 
 on the rescaled energy $\epsilon$ and on the system size $L$. 
 To answer the question about energy dependence, we consider two representative values of the rescaled energy 
 $\epsilon=0.2, 0.5$. The system size dependence of $\beta_0$ is probed by studying Heisenberg spin chains
 of length $L=14,16,18,20$.

 \begin{figure}[t]
\includegraphics[width=0.98\linewidth]{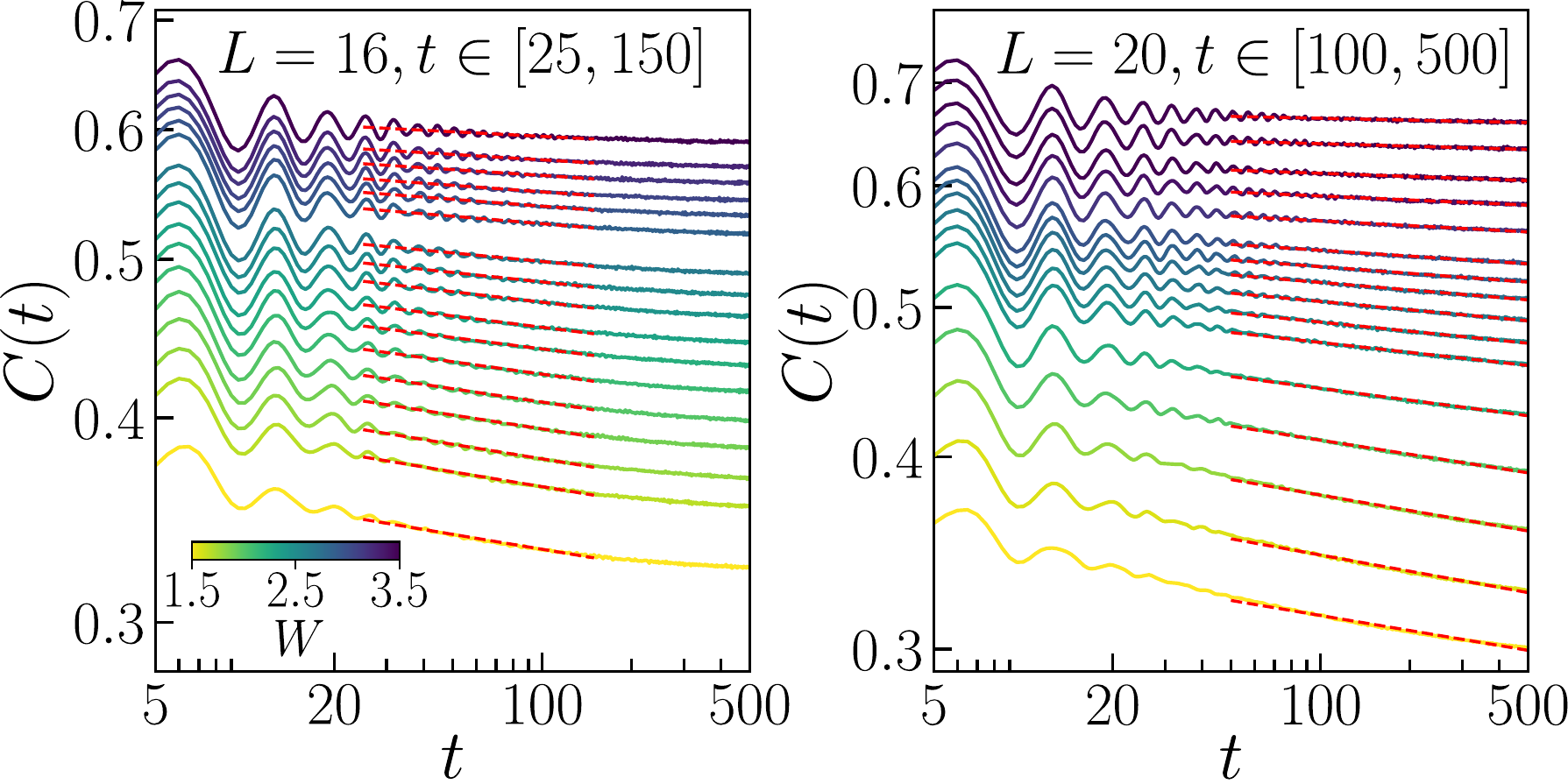}
  \caption{Density correlation function $C(t)$ for system sizes $L=16, 20$ -- left and right panel, respectively; 
  rescaled energy of the initial state $\epsilon=0.2$, disorder strength $W=1.5, ..., 3.5$ (color coded). Data are averaged over more
  than $5000$ disorder realizations. Power-law fits $C(t)\propto t^{-\beta}$ in time intervals denoted on respective panels
  are denoted by the dashed lines.
  }
  \label{ctL14_20_E02}
\end{figure}

Using eigenvalues $E_i$ obtained in the exact diagonalization,
we calculate the gap ratios $r_i = \min \{ \frac{s_{i+1}}{s_i}, \frac{s_i}{s_{i+1}} \}$,
where $s_i=E_{i+1}-E_i$, and average it over small fraction of system spectrum around the rescaled
energy $\epsilon$ and over more than $2000$ disorder realizations obtaining an average gap ratio $\overline r$, 
shown in Fig.~\ref{rbars0205}.  The average gap ratio changes between $\overline r\approx0.53$ in ergodic 
phase of systems with time reversal invariance and $\overline r\approx0.39$ in MBL phase 
\cite{Oganesyan07}. In the calculations of the average gap ratio $\overline r$ we use
periodic boundary conditions to probe the bulk properties of the system that are also
reflected by the density correlation function $C(t)$ (for which we disregard the two sites
at the boundary of the system).

As an approximation for critical disorder strength at given system size $L$,
we use, following \cite{Sierant20c},
disorder strength $W_C(L)$ at which the curves $\overline r(W)$ for $L$ and $L+2$ cross. 
For the rescaled energy $\epsilon=0.2$ we obtain $W^{\epsilon=0.2}_C(L) \approx 2.7$ for all the system
sizes considered. Drift of the crossing points is much more 
pronounced for the rescaled energy $\epsilon=0.5$ for which we obtain $W^{\epsilon=0.5}_C(L=14) \approx 3.3$ 
which increases with system size $L$  up to $W^{\epsilon=0.5}_C(L=18) \approx 3.85$.

\begin{figure}[ht]
\includegraphics[width=0.98\linewidth]{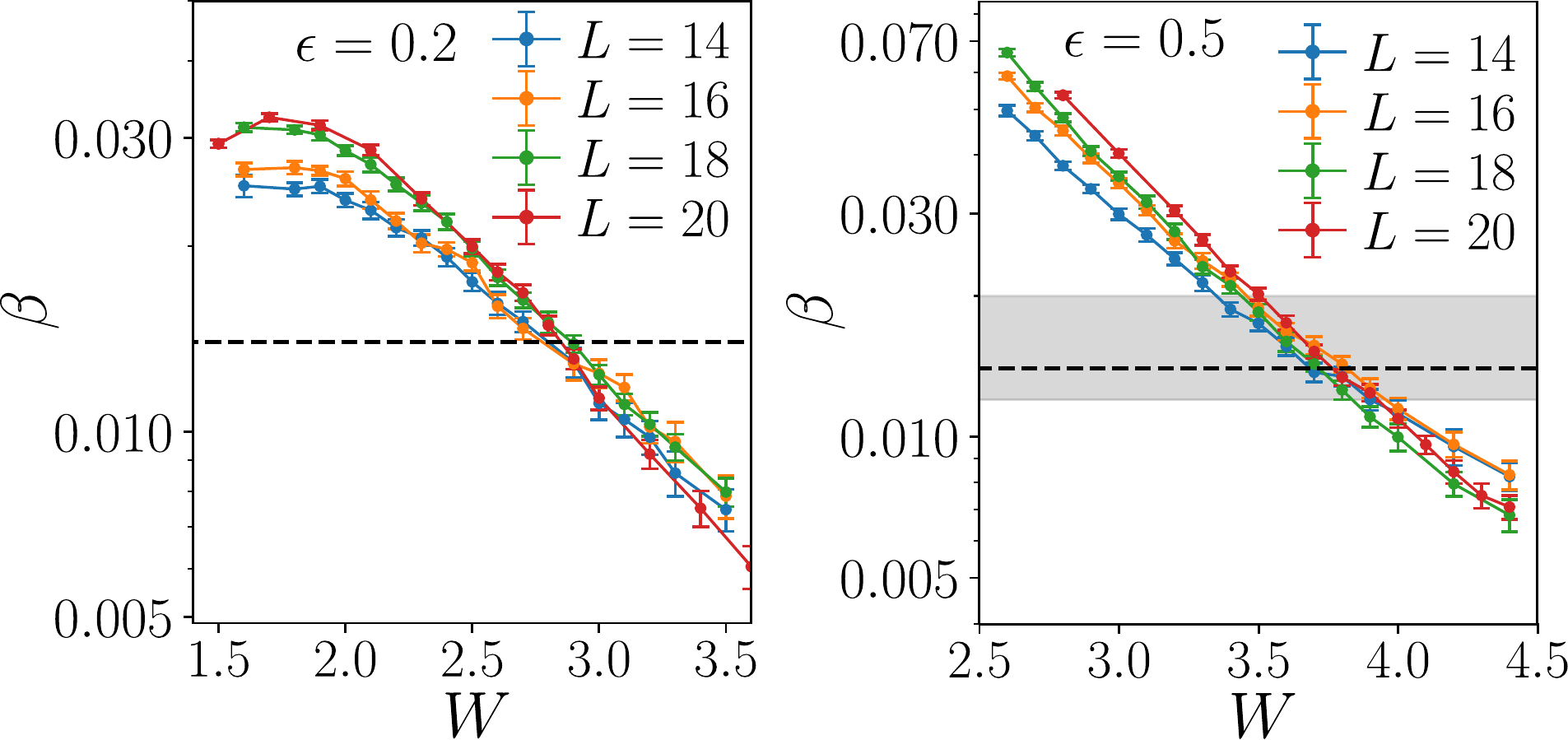}
  \caption{The exponent $\beta$, obtained in fitting the density correlation
  function $C(t)$ with an algebraic decay $a_0 t^{-\beta}$ is
  plotted as function of disorder strength $W$ for the rescaled energy $\epsilon=0.2$ (left panel)
  and $\epsilon=0.5$ (right panel).
  The errorbars represent the $1\sigma$ errors of the fitting obtained from statistical resampling of disorder
  realizations.
 The dashed line shows the cut-off exponent $\beta_0 = 0.014$.
  }
  \label{betalogE02E05}
\end{figure}  
The obtained approximations of critical disorder strength $W^{\epsilon=0.5}_C(L)$ can now be used to 
find the corresponding values of cut-off $\beta_0$ at system size $L$. To that end, we calculate
density correlations functions (examples shown in Fig.~\ref{ctL14_20_E02}) for $14\leqslant L \leqslant20$. 
Exponents $\beta$ are obtained by power-law fits to $C(t)$. At small system sizes, $C(t)$ tend to saturate at times of
the order of few hundred (in units of $J^{-1}$) that roughly correspond to inverse level spacing at the
rescaled energy $\epsilon$. Thus, we vary the interval in which fitting is performed with system size: $t\in[20,100]$
for $L=14$; $t\in[25,150]$ for $L=16$; $t\in[50,300]$ for $L=18$ and $t\in[100,500]$ for $L=20$.
The resulting values of $\beta$ are shown in Fig.~\ref{betalogE02E05}. For $\epsilon=0.2$, we
observe that $W^{\epsilon=0.2}_C \approx 2.7$ is consistent with the cut-off $\beta_0=0.014(2)$, confirming that
the cut-off $\beta_0=0.014$ can be used globally, for all considered values of the rescaled energy.
For $\epsilon=0.5$, the drift of the crossing point $W^{\epsilon=0.5}_C(L=14) \approx 3.3$ to
$W^{\epsilon=0.5}_C(L=18) \approx 3.85$ corresponds to change in value of the 
$\beta_0$ between $0.02$ and $0.012$ -- as denoted by the shaded area in Fig.~\ref{betalogE02E05}.
This decreasing trend of $\beta_0$ with system size $L$ at $\epsilon=0.5$ might suggest that the cut-off
value decreases with system size. Within the current approach we are unable to conclude whether $\beta_0$ decreasing
with system size is a better hypothesis than the constant value of $\beta_0$.
This illustrates the difficulties inherent to studying MBL transition in thermodynamic limit, described, e.g. in 
\cite{Panda19}.
Nevertheless, as Fig. 7 of the main text shows, changes of the cut-off $\beta_0$ in range from $0.01$ to $0.02$ do not
affect our conclusions about the existence or the shape of the many-body mobility edge in the system.

\bibliographystyle{apsrev4-1}
\bibliography{ref_05_20.bib} 
%\bibliography{mobil_v3.bbl} 
%\input{v7.bbl}
%\include{supplement}